\documentclass[conference]{IEEEtran}
\IEEEoverridecommandlockouts

\usepackage{cite}
\usepackage{amsmath,amssymb,amsfonts}
\usepackage{algorithmic}
\usepackage{graphicx}
\usepackage{textcomp}
\usepackage{xcolor}
\usepackage{booktabs}  
\usepackage{multirow}  
\usepackage{adjustbox} 
\usepackage{algorithm}
\usepackage{subcaption}
\usepackage{makecell}
\usepackage{amsmath}
\usepackage{enumitem}
\usepackage{lettrine} 
\usepackage{hyperref} 
\usepackage[absolute,overlay]{textpos}
\hypersetup{
    colorlinks=true,
    linkcolor=black,
    filecolor=magenta,      
    urlcolor=black,
    citecolor=blue,
}

\usepackage{caption}
\def\BibTeX{{\rm B\kern-.05em{\sc i\kern-.025em b}\kern-.08em
    T\kern-.1667em\lower.7ex\hbox{E}\kern-.125emX}}
\begin{document}

\title{
FastMamba: A High-Speed and Efficient Mamba Accelerator on FPGA with Accurate Quantization
}

\author{
\IEEEauthorblockN{Aotao Wang$^{1}$, Haikuo Shao$^{1}$, Shaobo Ma$^{1}$, and Zhongfeng Wang$^{1,2}$}
\IEEEauthorblockA{$^{1}$School of Electronic Science and Engineering, Nanjing University, Nanjing, China \\
$^{2}$School of Integrated Circuits, Sun Yat-sen University, Shenzhen, China \\
Email: \{atwang, hkshao, shaoboma\}@smail.nju.edu.cn, zfwang@nju.edu.cn}

\thanks{This work was supported in part by the National Key R\&D Program of China under Grant 2022YFB4400600, and in part by the Postgraduate Research \& Practice Innovation Program of Jiangsu Province under Grant KYCX24\textunderscore0149. \textit{(Corresponding author: Zhongfeng Wang.)}}
}

\maketitle

\begin{textblock*}{8.9cm}(1.36cm,25.5cm)
  {\fontsize{8}{10}\selectfont
  979-8-3315-3477-6/25/\$31.00~\copyright2025 IEEE}
\end{textblock*}  

\begin{abstract}
State Space Models (SSMs), like recent Mamba2, have achieved remarkable performance and received extensive attention.
However, deploying Mamba2 on resource-constrained edge devices encounters many problems: severe outliers within the linear layer challenging the quantization, diverse and irregular element-wise tensor operations, and hardware-unfriendly nonlinear functions in the SSM block. 
To address these issues, this paper presents FastMamba, a dedicated accelerator on FPGA with hardware-algorithm co-design to promote the deployment efficiency of Mamba2.
Specifically, we successfully achieve 8-bit quantization for linear layers through Hadamard transformation to eliminate outliers. Moreover, a hardware-friendly and fine-grained power-of-two quantization framework is presented for the SSM block and convolution layer, and a first-order linear approximation is developed to optimize the nonlinear functions. 
Based on the accurate algorithm quantization, we propose an accelerator that integrates parallel vector processing units, pipelined execution dataflow, and an efficient SSM Nonlinear Approximation Unit, which enhances computational efficiency and reduces hardware complexity. 
Finally, we evaluate FastMamba on Xilinx VC709 FPGA. 
For the input prefill task on Mamba2-130M, FastMamba achieves 68.80$\times$ and 8.90$\times$ speedup over Intel Xeon 4210R CPU and NVIDIA RTX 3090 GPU, respectively. 
In the output decode experiment with Mamba2-2.7B, FastMamba attains 1.65$\times$ higher energy efficiency than RTX 3090 GPU. 
\end{abstract}


\section{Introduction}


Mamba\cite{b1}, as a novel class of State Space Model (SSM), has become a prominent foundation model architecture and has been applied in various deep learning fields such as natural language processing\cite{b1,b2}, images analysis\cite{b3,b4}, and video processing\cite{b5}. It effectively addresses the quadratic growth in computational complexity of traditional Transformer models \cite{b6} concerning input sequence length, offering a more efficient solution for processing longer sequences. Recent Mamba2\cite{b2} has further reduced computational complexity while improving accuracy through state space duality (SSD) framework and semiseparable matrix decompositions and is 2$\sim$8$\times$ faster than Mamba. 


Deploying models on personal edge devices, rather than cloud servers like the Graphics Processing Unit (GPU), has attracted continuous attention for real-time processing and data privacy considerations.
Despite Mamba2's advantages, running it on edge hardware like the Field-Programmable Gate Array (FPGA) still faces significant design challenges due to its rigorous computation and memory requirements.
Existing algorithm optimization methods, such as quantization \cite{b7},\cite{b8}, are primarily focused on large models based on Transformer architecture, with limited exploration in the context of Mamba.
Previously dedicated Mamba ASIC accelerator\cite{b9} also lacks sufficient application of quantization methods, which limits its hardware efficiency and flexibility.

The Mamba2 architecture consists of the linear layer, convolution layer, SSM block, and nonlinear functions including Root Mean Square (RMS) normalization \cite{b10} and $SiLU$ activation \cite{b11}.
To identify performance bottlenecks, we profiled the runtime of each component in the prompt prefill experiment of Mamba2-130M on RTX 3090 GPU.
As shown in Fig. \ref{fig:runtime}, the SSM block and linear layer occupy a majority of the computational load.
The runtime proportion of the SSM also increases with the growth of input sequence length.
Various kinds of element-wise tensor operations (e.g. add, multiplication, and nonlinear functions) and massive matrix multiplications between activation values and weights \cite{b13}, constitute the primary computations in the SSM block and the linear layer, respectively.
For edge deployment of Mamba2, the dominant SSM block and linear layer face the following challenges:
\begin{figure}[tbp]
\centerline{\includegraphics[width=0.5\textwidth]{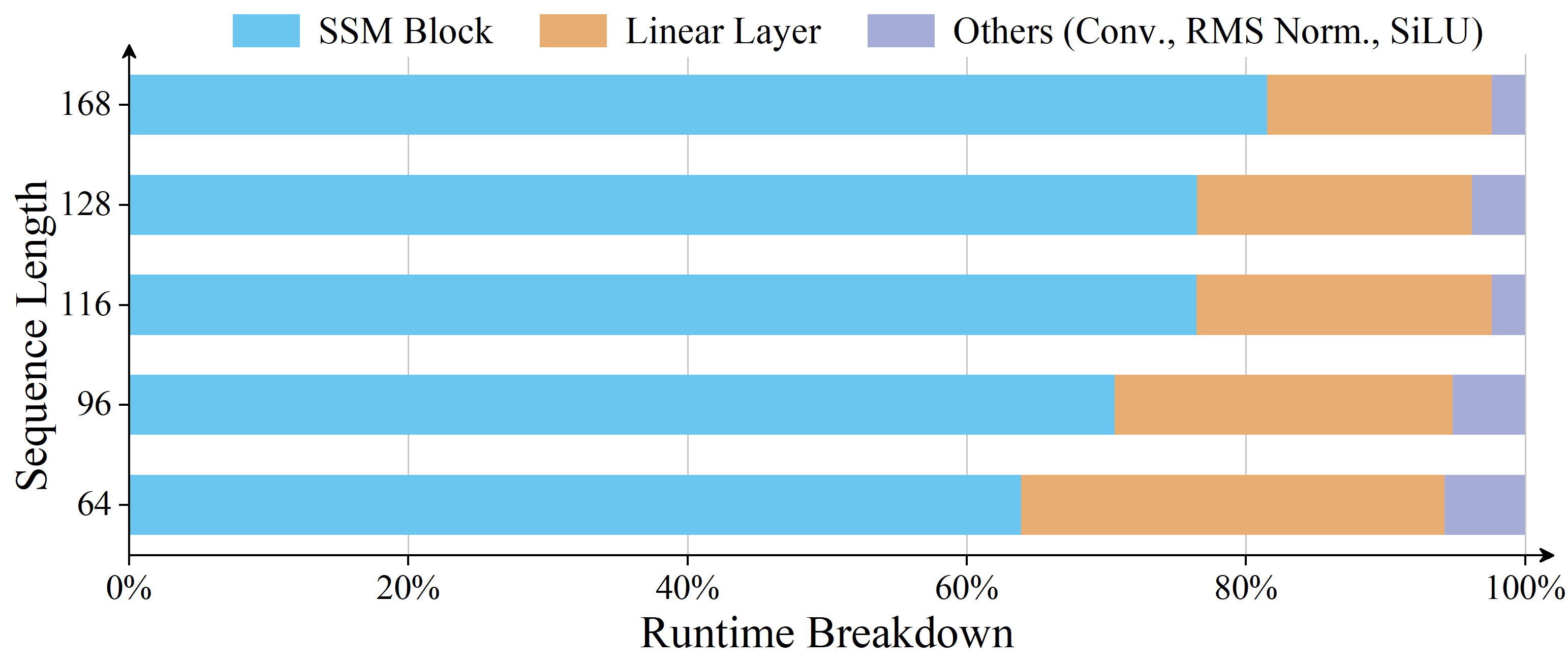}}
\caption{Runtime breakdown with different evaluation sequence lengths during the prompt prefill stage.}
\vspace{-2.5em}
\label{fig:runtime}
\end{figure}
(1) The activation values and weights in the linear layer exhibit severe outlier distributions. Conventional quantization methods fail to mitigate the impact of outlier characteristics, resulting in significant accuracy degradation.
Moreover, the nonlinear operations used in SSM block, such as the exponential and $SoftPlus$ \cite{b14} functions, are sensitive to quantization, necessitating carefully designed schemes to maintain accuracy.
(2) Diverse and irregular element-wise operations in the SSM block constitute the majority of computational load, presenting the bottleneck to acceleration. So it's necessary to design an efficient hardware architecture to unify various computations. The hardware-unfriendly nonlinear functions inside the SSM block also increase hardware design complexities.


In this paper, we propose FastMamba, an efficient Mamba accelerator with algorithm and hardware co-design to address the above challenges. Our main contributions are as follows:

\textbf{(1) Accurate quantization of the linear layer, convolution layer, and SSM block.}
We successfully quantize linear layers to 8-bit through Hadamard transformation\cite{b8}, which eliminates outlier distributions in the linear layer. A hardware-friendly and fine-grained power-of-two (PoT) quantization framework is presented for the convolution layer and SSM block. The quantized algorithm significantly reduces the computational complexity with accuracy degradation within $1\%$.

\textbf{(2) Efficient accelerator design.} We present parallel vector processing units (VPUs) executing basic computations in Mamba2. Based on VPUs, we design efficient hardware for the SSM block and linear layer through group parallelism and pipelined design.
Moreover, within the SSM block, we unify $SoftPlus$ and exponential functions into hardware-friendly and efficient linear approximation and further design a multi-mode dedicated unit, which effectively saves Digital Signal Processor (DSP) and flip-flop resources.

\textbf{(3) Experimental evaluations.} The accelerator is implemented on the Xilinx Virtex-7 VC709 FPGA. For the prompt prefill task on Mamba2-130M, FastMamba achieves a maximum speedup of 68.80$\times$ and 8.90$\times$ compared to the Intel Xeon 4210R ‌Central Processing Unit (CPU) and NVIDIA RTX 3090 GPU, respectively. In the decode experiment with Mamba2-2.7B, the design achieves a 1.65$\times$ improvement in energy efficiency compared to the RTX 3090 GPU.

\section{Background}
\subsection{State Space Model}
The SSM is a mathematical framework that uses hidden parameters, called “states,” to capture temporal correlations and data relationships. The classical continuous-time SSM, as shown in Eq. \eqref{SSMCfunction}, consists of a state equation and an output equation. These equations describe the mapping from the input signal $x(t)\in \mathbb{R}$ to the output signal $y(t)\in \mathbb{R}$ at time $t$, mediated by the hidden state $h(t)\in \mathbb{R}^N$:
\begin{equation}
\begin{aligned}
h'(t) &= Ah(t) + Bx(t)\\
y(t) &= Ch(t) + Dx(t)
\label{SSMCfunction}
\end{aligned}
\end{equation}
here, $h'(t)\in \mathbb{R}^N$ is the time derivative of $h(t)\in \mathbb{R}^N$,  indicating how $h(t)$ changes over time. The key parameters are the state-transition matrix $A\in \mathbb{R}^{N\times N}$, the input matrix $B\in \mathbb{R}^{N\times 1}$, the output matrix $C\in \mathbb{R}^{N\times 1}$, and the feedthrough matrix $D\in \mathbb{R}$. 

In practical machine-learning scenarios, data is usually discrete. Thus, discretizing the continuous-time Eq. \eqref{SSMCfunction} is essential. Discretization aims to convert continuous-time parameters like $A$ and $B$ into discrete-time parameters. The Zero-Order Hold (ZOH) method \cite{b15} is a common approach in the discretization process. It introduces $\Delta$, the time interval during which the function value is held constant. Applying the ZOH method transforms the continuous-time Eq. \eqref{SSMCfunction} into the following discrete-time Eq. \eqref{SSMDfunction}:
\begin{equation}
\begin{aligned}
h_k &= \overline{A}h_{k - 1}+\overline{B}x_k \\
y_k &= Ch_k + Dx_k
\label{SSMDfunction}
\end{aligned}
\end{equation}
where $\overline{A} = \exp(\Delta A)$,  $\overline{B} = (\Delta A)^{-1}(\exp(\Delta A) - I)\cdot\Delta B \approx \Delta B$, and  $\Delta$ is the discretization step size. In a discrete-time system, $\overline{A}$ and $\overline{B}$ are calculated based on $\Delta$.

\subsection{Mamba}
\begin{figure}[tbp]
\centerline{\includegraphics[width=0.5\textwidth]{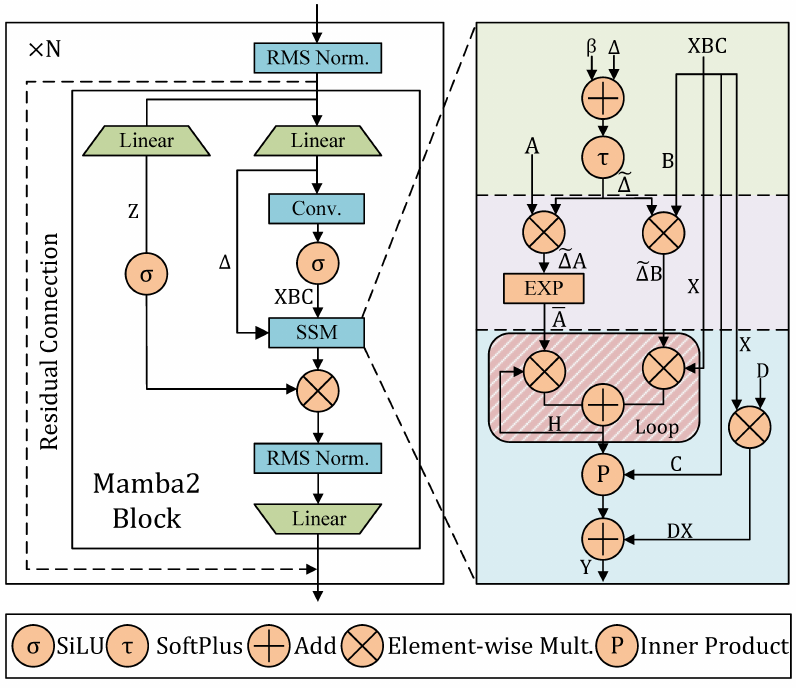}}
\caption{Computational flow in Mamba2 block and SSM block.}
\vspace{-2.2em}
\label{dataflow}
\end{figure}

The overall computational process of Mamba2 is depicted in Fig. \ref{dataflow}. Each layer in Mamba2 consists of 2 RMS normalization layers, multiple linear layers, a convolution layer, 2 $SiLU$ (denoted as $\sigma$), an SSM block, and a residual connection. 

The input to the SSM block is the data from the $Silu$, split into $X$, $B$, and $C$ along the feature dimension and the $\Delta$ from the linear layer.
The computational flow of the SSM block is partitioned for hardware optimization. 
Firstly, $\Delta$ undergoes a preprocessing operation to generate $\widetilde{\Delta}$.
Then, $\widetilde{\Delta}$ is processed with $A$ and $B$, producing $\overline{A}$ and $\widetilde{\Delta}B$ for subsequent iterations.
Finally, after $L$ (sequence length) iterations, the current $L$-time $H$ performs the inner product operation with $C$, and the result is added to the bypass result $DX$ to produce the final output.

In a real-world deployment, Mamba2 involves two key processes: prompt prefill and token decode. Prompt prefill requires computing with the entire input sequence as context, resulting in intensive GPU memory consumption. In contrast, token decode generates one token by one token. This characteristic enables the decode stage to support inference of large-scale Mamba2 models even with limited hardware resources.

\section{Algorithm Quantization}
\subsection{Hadamard-based Linear Quantization}
\begin{figure}[tbp]
\centerline{\includegraphics[width=0.5\textwidth]{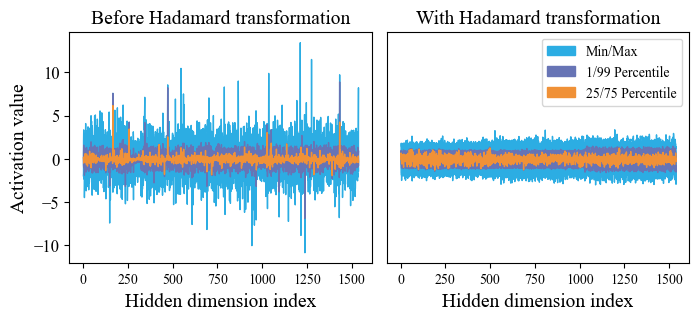}}
\caption{The distributions of activation values before Hadamard transforming and with Hadamard transforming.}
\label{activation}
\vspace{-1.8em}
\end{figure}

In Mamba2, activation values in linear layers typically exhibit more extreme outlier distributions than weights, making their quantization much more difficult. 
The impact of outliers in activation values $X$ and weights $W$ can be weakened by the Hadamard quantization method \cite{b8} based on the equation of $nY=(XH)(H^TW^T)$, which Hadamard matrix $H\in R^{n\times n}$ and satisfies $n=2^k(k\in\textbf{Z})$. Specifically, when normalized by $\sqrt{n}$, the Hadamard matrix becomes orthonormal. Therefore, Hadamard transformation is essentially an orthonormal transformation.
As shown in Fig. \ref{activation}, the activation values exhibit a more concentrated distribution with a narrower dynamic range following the Hadamard transformation.

This transformation improves the stability of low-bit quantization. The proposed Hadamard-based Linear Quantization Method (Algorithm \ref{alg:hadamard_quant}) uses the special properties of the Hadamard transformation for the 8-bit quantization of activation values and weights. 

The activation matrix $\textbf{X}$ and the weight matrix $\textbf{W}$ are uniformly partitioned into $m$ groups of sub-matrices $\textbf{X}[i]$ and $\textbf{W}[i]$ to ensure dimension $\frac{d}{m}=2^k(k\in\textbf{Z})$. The Hadamard matrix $\textbf{H}[i]$ is constructed according to the dimension of $\textbf{X}[i]$. 
Subsequently, the scaling factors $s_X$ and $s_W$ are computed based on the maximum values of $\textbf{X}_H$ and $\textbf{W}_H$, and these scaling factors are then utilized to generate the 8-bit $\hat{\textbf{X}}_H[i]$ and $\hat{\textbf{W}}_H[i]$ within the range of $-128\sim127$. 
Ultimately, the partial sums of each group $\hat{\textbf{Y}}$ are reduced, and the result is de-quantized to output the result \textbf{Y} of the linear layer.

\subsection{SSM Quantization and Nonlinear Approximation}
PoT quantization represents a hardware-efficient method for converting floating-point to fixed-point by applying a scaling factor of $2^p(p\in\textbf{Z})$.
As illustrated in the right section of Fig. \ref{dataflow}, the linear operations like add, element-wise multiplication, and inner products are quantified by PoT. 
However, the PoT quantization can't be directly applied to the nonlinear operations including $SoftPlus$ activation and exponential function. 
Therefore, we optimize these nonlinear functions within a first-order linear approximation algorithm framework aimed at supporting fixed-point quantization by PoT and improving computing efficiency.
\begin{figure}[!t]
\begin{algorithm}[H]
\renewcommand{\algorithmicrequire}{\textbf{Input:}}
\renewcommand{\algorithmicensure}{\textbf{Output:}}
\caption{Hadamard-based Linear Quantization Method} 
\label{alg:hadamard_quant}
\begin{algorithmic}[1]
\REQUIRE $\textbf{X}\in \mathbb{R}^{l\times d},\textbf{W}\in \mathbb{R}^{q\times d}$
\ENSURE $\textbf{Y}\in \mathbb{R}^{l\times q}$
\STATE{$\textbf{X}[i]\in \mathbb{R}^{l\times \frac{d}{m}}\leftarrow \textbf{X}\in \mathbb{R}^{l\times d},i=0,1,\ldots,m-1$}
\STATE{$\textbf{W}[i]\in \mathbb{R}^{q\times \frac{d}{m}}\leftarrow \textbf{W}\in \mathbb{R}^{q\times d},i=0,1,\ldots,m-1$}
\STATE{$\textbf{for }i=0,1,\ldots,m-1\textbf{ do}$}
\STATE{$\quad \textbf{H}[i]\in \mathbb{R}^{\frac{d}{m}\times\frac{d}{m}}\leftarrow \text{FindHadamard}(\textbf{X}[i])$}
\STATE{$\quad \textbf{X}_H[i] = \textbf{X}[i]\textbf{H}[i]$}
\STATE{$\quad\textbf{W}_H[i] = \textbf{H}^T[i]\textbf{W}^T[i]$} 
\STATE{$s_X = \text{FindScale}(\text{cat}(\textbf{X}_H[0],\ldots,\textbf{X}_H[m-1]))$}
\STATE{$s_W = \text{FindScale}(\text{cat}(\textbf{W}_H[0],\ldots,\textbf{W}_H[m-1]))$}
\STATE{$\textbf{for }i=0,1,\ldots,m-1\textbf{ do}$}
\STATE{$\quad \hat{\textbf{X}}_H[i] = \text{Quant}(\textbf{X}_H[i],s_X)$}
\STATE{$\quad \hat{\textbf{W}}_H[i] = \text{Quant}(\textbf{W}_H[i],s_W)$}
\STATE{$\quad \hat{\textbf{Y}} = \hat{\textbf{Y}}+\hat{\textbf{X}}_H[i]\hat{\textbf{W}}_H[i]$}
\STATE{$\textbf{Y} = \hat{\textbf{Y}}\cdot s_X s_W \cdot \frac{m}{d}$}
\end{algorithmic} 
\end{algorithm}
\vspace{-2em}
\end{figure}
\subsubsection{Exponential linear approximaion}
Statistical analysis reveals that all values of the $\widetilde{\Delta}$ tensor are less than 0. Thus, we employ an exponential linear approximation algorithm tailored for the negative-number domain \cite{b16}. The algorithm formula is:
\begin{equation}
\begin{aligned}
e^{x} &= 2^{x\log_2 e},\quad x \leq 0, \log_2 e\approx {(1.0111)}_2 \\
      &= 2^{(u)+(v)} = 2^{u} \cdot 2^{v} = 2^{v} \gg |u|,~ u,v\leq 0.
\label{func:exp}
\end{aligned}
\end{equation}
Here, $u$ and $v$ are the integer and fractional part of $x\log_2 e$ respectively, and $\gg$ denotes a shift operation. Since $v\in (-1,0]$, we perform an 8-segment high-precision first-order linear approximation of $2^v$. 
\subsubsection{SoftPlus Symmetry Algorithm}
The $SoftPlus$ activation function has the following symmetric property \cite{b17}:
\begin{equation}
\begin{aligned}
SoftPlus(x) &= x + SoftPlus(-x)
\end{aligned}
\end{equation}
We approximate $ln(x)$ using the following first-order linear approximation:
\begin{equation}
\begin{aligned}
SoftPlus(x)=\ln(1 + e^{x})\approx e^{x}
\end{aligned}
\end{equation}
Combining Eq. \eqref{func:exp} and the symmetric property of $SoftPlus$, we transform $SoftPlus$ into the following system of equations: 
\begin{equation}
\begin{aligned}
SoftPlus(x)=
\begin{cases}
 e^{x}, & x \leq 0, \\
e^{-x} + x, & x > 0.
\end{cases}
\end{aligned}
\end{equation}
Through these transformations, for inputs greater than 0, the result of $SoftPlus$ can also be calculated using Eq. \eqref{func:exp}. Ultimately, the nonlinear functions in SSM are unified within the framework of first-order linear calculations. The hardware implementation of $SoftPlus$ can reuse the approximate unit of the exponential function, saving hardware resources and reducing computational complexity.

\section{Hardware Architecture}
\subsection{Architecture Overview}

\begin{figure}[tp]
\centerline{\includegraphics[width=0.5\textwidth]{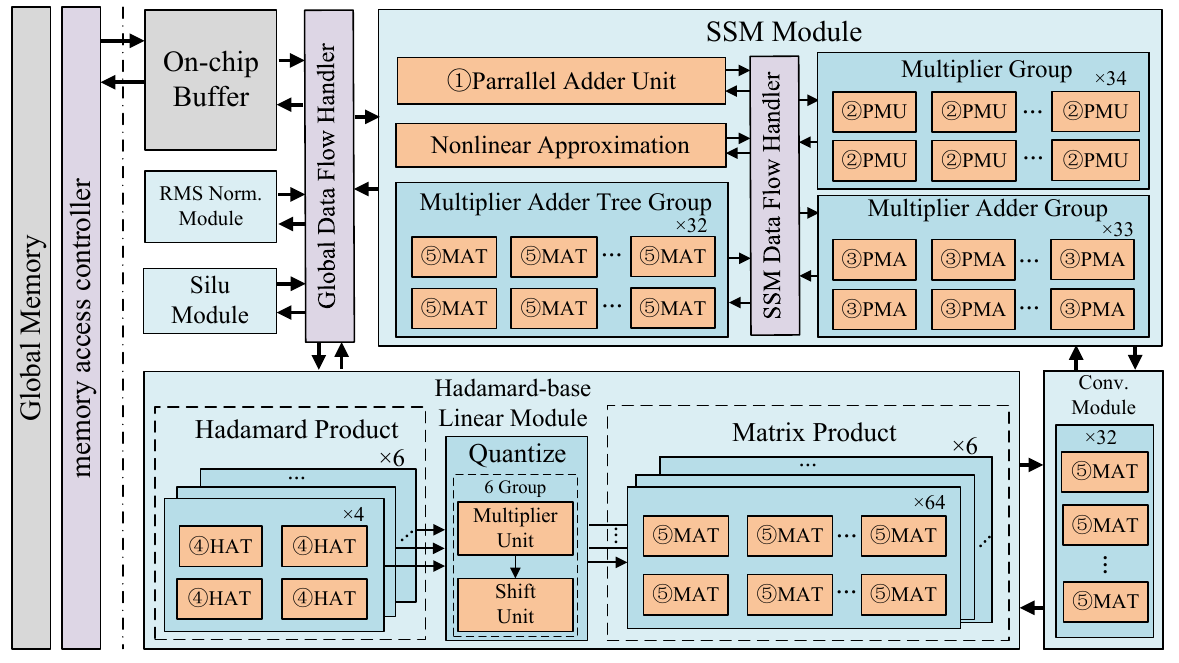}}
\caption{Overall Architecture of FastMamba.}
\label{fig:ArcM}
\vspace{-1.5em}
\end{figure}

The overall architecture of FastMamba is shown in Fig. \ref{fig:ArcM}, which consists of the fixed-point computing group (Hadamard-based Linear module, Convolution Module, and SSM Module), as well as the floating-point computing group (RMS Normalization Module and SiLU Module). 
The data is loaded from Global Memory into the On-chip Buffer.
The dataflow between functional modules and buffers is managed by the Data Flow Handler.

\begin{figure}[tp]
\centerline{\includegraphics[width=0.5\textwidth]{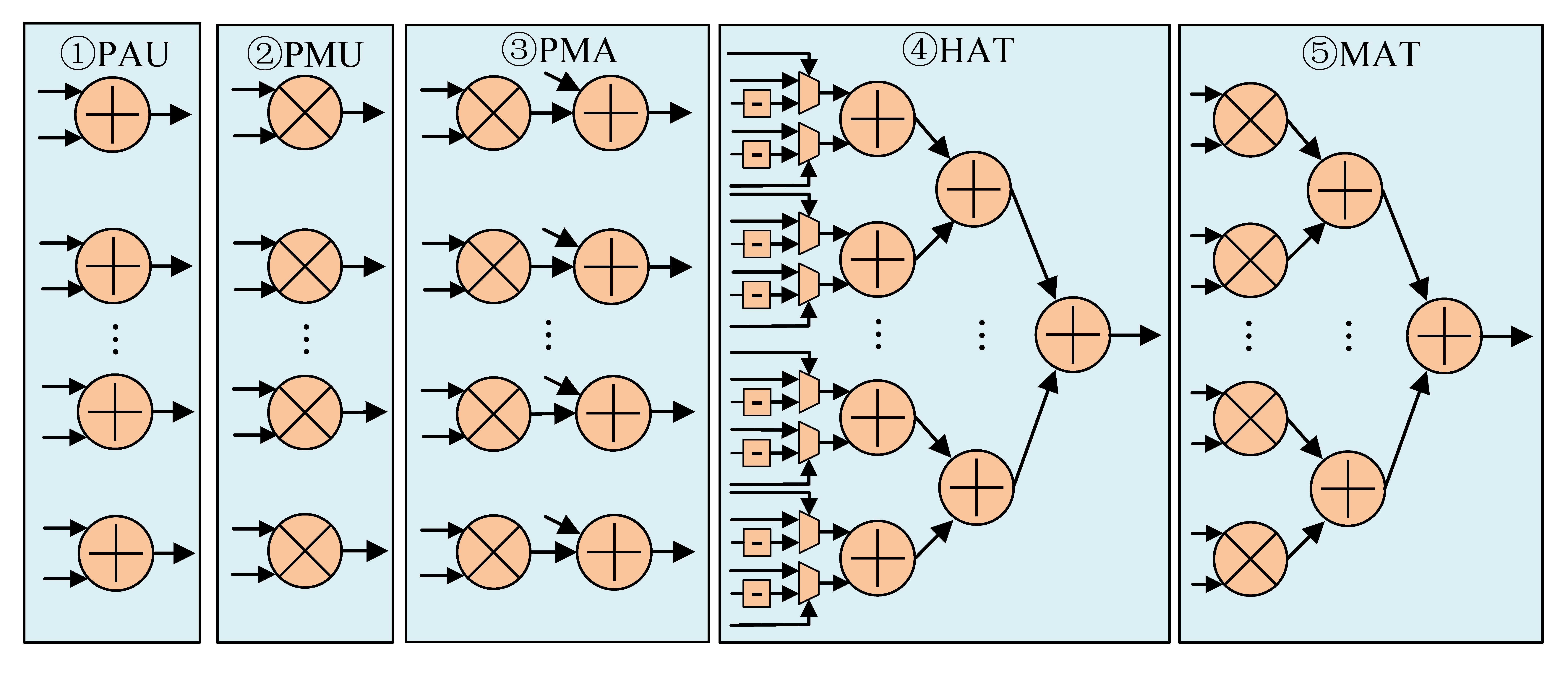}}
\caption{The structure of the multipliers and adders in the VPUs.}
\label{fig:VPU}
\vspace{-1.5em}
\end{figure}

In FastMamba, computations can be categorized into addition, multiplication, multiply-add, accumulation, and multiply-accumulate.
To handle these operations, we propose VPUs consisting of multipliers and adders as presented in Fig. \ref{fig:VPU}, which involve five types: \textcircled{\scriptsize{1}} Parallel Adder Unit (PAU), \textcircled{\scriptsize{2}} Parallel Multiplier Unit (PMU), \textcircled{\scriptsize{3}} Parallel Multiplier Adder Unit (PMA), \textcircled{\scriptsize{4}} Hadamard Adder Tree (HAT), and \textcircled{\scriptsize{5}} Multiplier Adder Tree (MAT).
The outputs of PAU, PMU, and PMA are vectors of the same length as the inputs, while the outputs of HAT and MAT are reduced to scalars.
The inputs, outputs, and corresponding computational functions of each VPU are listed in Table \ref{tab:VPUF}.

\begin{table}[htbp]
\centering
\caption{Function Configuration for VPUs}  
\label{tab:VPUF}
\begin{adjustbox}{width=0.5\textwidth}
\Large 
\begin{tabular}{ccccccc}  
\Xhline{1pt}
\addlinespace 
\textbf{Number} & \textbf{VPU} & \textbf{Input0:Length} & \textbf{Input1:Length} & \textbf{Input2:Length} & \textbf{Output:Length} & \textbf{Funciton} \\
\addlinespace 
\hline  
\addlinespace 
\textcircled{\small{1}} & PAU & $A:n$ & $B:n$ & $-$ & $P:n$ & $A+B=P$ \\
\addlinespace 
\hline  
\addlinespace 
\textcircled{\small{2}} & PMU & $A:n$ & $B:n$ & $-$ & $P:n$ & $A\times B=P$ \\
\addlinespace 
\hline  
\addlinespace 
\textcircled{\small{3}} & PMA & $A:n$ & $B:n$ & $C:n$ & $P:n$ & $A\times B+C=P$ \\
\addlinespace 
\hline  
\addlinespace 
\textcircled{\small{4}} & HAT & $A:n$ & $-$ & $-$ & $P:1$ & $\sum_{i = 1}^{n}(A_i)=P$ \\
\addlinespace 
\hline  
\addlinespace 
\textcircled{\small{5}} & MAT & $A:n$ & $B:n$ & $-$ & $P:1$ & $\sum_{i = 1}^{n}(A_i\times B_i)=P$ \\
\addlinespace 
\Xhline{1pt}
\end{tabular}
\end{adjustbox}
\vspace{-1em}
\end{table}

Based on VPUs, the overview of functional modules in different regions of FastMamba is as follows:

\textbf{Hadamard-base Linear Module:} This module is designed with 6 parallel computing groups. Each group is designed to perform the Hadamard product, quantization, and 8-bit matrix multiplication.

\textbf{Convolution Module:} This module incorporates 32 MAT units to perform 1-D convolution. For a 1-D convolution with a kernel size of 4, each MAT performs the convolution operation on a vector length of 4.

\textbf{SSM Module:} Various VPUs are employed in this module to perform fixed-point computations required by SSM block. A Nonlinear Approximation Unit is also designed to compute exponential and $SoftPlus$ functions with 16-bit fixed-point data based on our algorithm approximation. 

\textbf{Floating-point Modules:} As demonstrated in Fig. \ref{fig:runtime}, the RMS normalization and $SiLU$ activation contribute a relatively small proportion in Mamba2 computational load. Utilizing floating-point computing units can effectively avoid accuracy loss with slight hardware overhead.


\begin{figure}[!t]
\centerline{\includegraphics[width=0.5\textwidth]{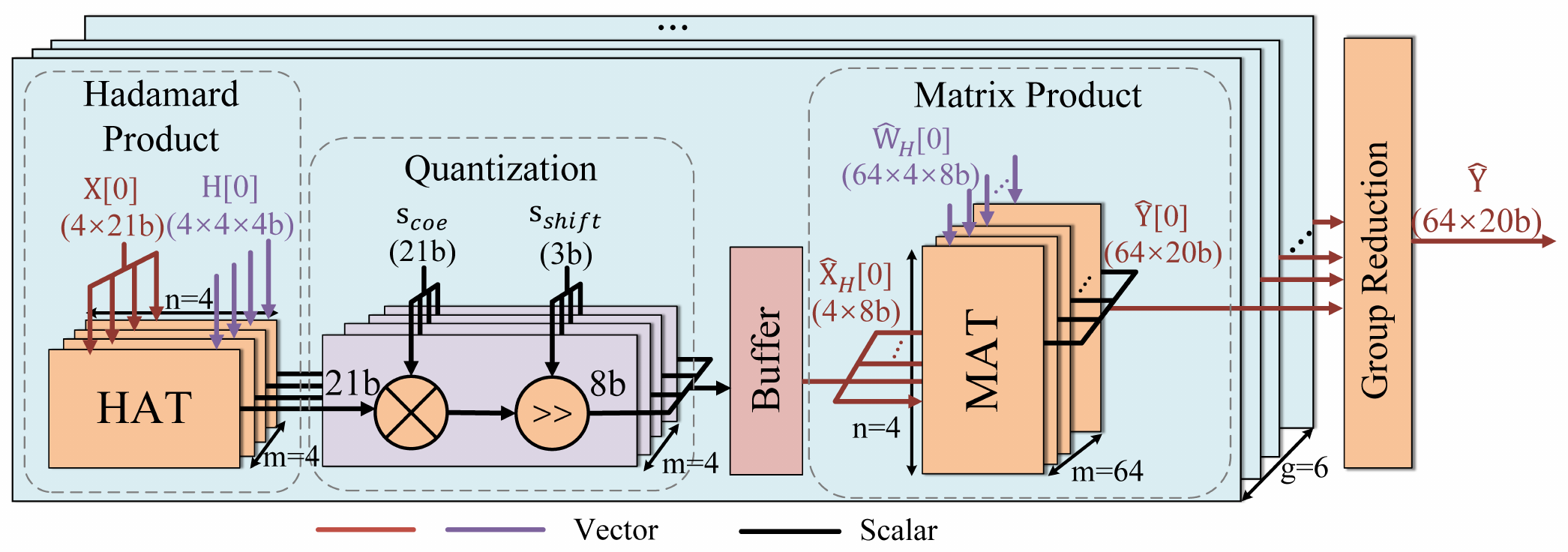}}
\caption{The architecture of Hadamard-base Linear Module (4$\times$21b indicates 4 21-bit data).}
\label{fig:linear}
\vspace{-1.8em}
\end{figure}

\begin{table*}[!t]
\centering
\caption{Comparison of Perplexity and Accuracy of different quantization algorithms on Mamba2-130M.}  
\label{tab:accruacy}
\begin{adjustbox}{width=0.9\textwidth}
\LARGE
\begin{tabular}{ccccccccccc}  
\toprule  
 \multirow{2}{*}{Method}& \multirow{2}{*}{Precision}&Lambada&Lambada&Hellaswag&Piqa&Arc-easy&Arc-challenge&Winogrande&Openbookqa&Average \\  
 &  &PPL($\downarrow$) & ACC($\uparrow$) & ACC($\uparrow$) & ACC($\uparrow$) & ACC($\uparrow$) & ACC($\uparrow$) & ACC($\uparrow$) & ACC($\uparrow$) & ACC($\uparrow$)  \\  
\midrule  
NormalQ & W8A8 & 33.7 & 32.5 & 33.9 & 62.3 & 44.5 & 23.6 & 50.9 & 30.6 & 39.8 \\  
SmoothQ \cite{b7} & W8A8 & 19.1 & 42.9 & 34.8 & 64.1 & 46.6 & 23.8 & 52.0 & 27.6 & 41.7 \\
\textbf{FastMamba-LQ} & W8A8 &17.2&43.1&35.3&64.8 &47.2&24.5&53.1&30.4&\textbf{42.6} \\  
\textbf{FastMamba}& W8A8 & 17.9& 43.0 & 34.8 & 64.7 & 47.0 & 23.6 & 52.6 & 29.6 & \underline{42.2} \\  
Mamba2-130M & FP16 &16.9 & 43.9 & 35.3 & 64.9 & 47.4 & 24.2 & 52.1 & 30.6 & \textbf{42.6} \\  
\bottomrule  
\end{tabular}
\end{adjustbox}
\vspace{-1.5em}
\end{table*}

\subsection{Hadamard-based Linear Module}

The Hadamard-based Linear Module is designed with 6 computing groups (as shown in Fig. \ref{fig:linear}), where each group indexed by $i=0,\dots,5$, can fully execute the linear quantization computation and output partial sums. 

In the Hadamard product, 4 parallel HAT units share the input activation values $X[i]$ and respectively receive the Hadamard matrix $H[i]$ to compute and output 4 scalar intermediate values. These intermediate values go through multiplication ($\times s_{coe}$) and shifting ($\gg s_{shift}$) operations for hardware-friendly quantization, and are then spliced along the feature dimension to form an 8-bit quantized activation vector $\hat{X}_H[i]$ of length 4, which is temporarily stored in a buffer. 

This module also has 64 parallel MAT units for the matrix product. These units share the input $\hat{X}_H[i]$ and respectively receive 8-bit weights $\hat{W}_H[i]$ to output partial sums $\hat{Y}[i]$ of length 64. The partial sums of each group are reduced to generate the linear output $\hat{Y}$.

\subsection{SSM Module}
\begin{figure}[tp]
\centerline{\includegraphics[width=0.5\textwidth]{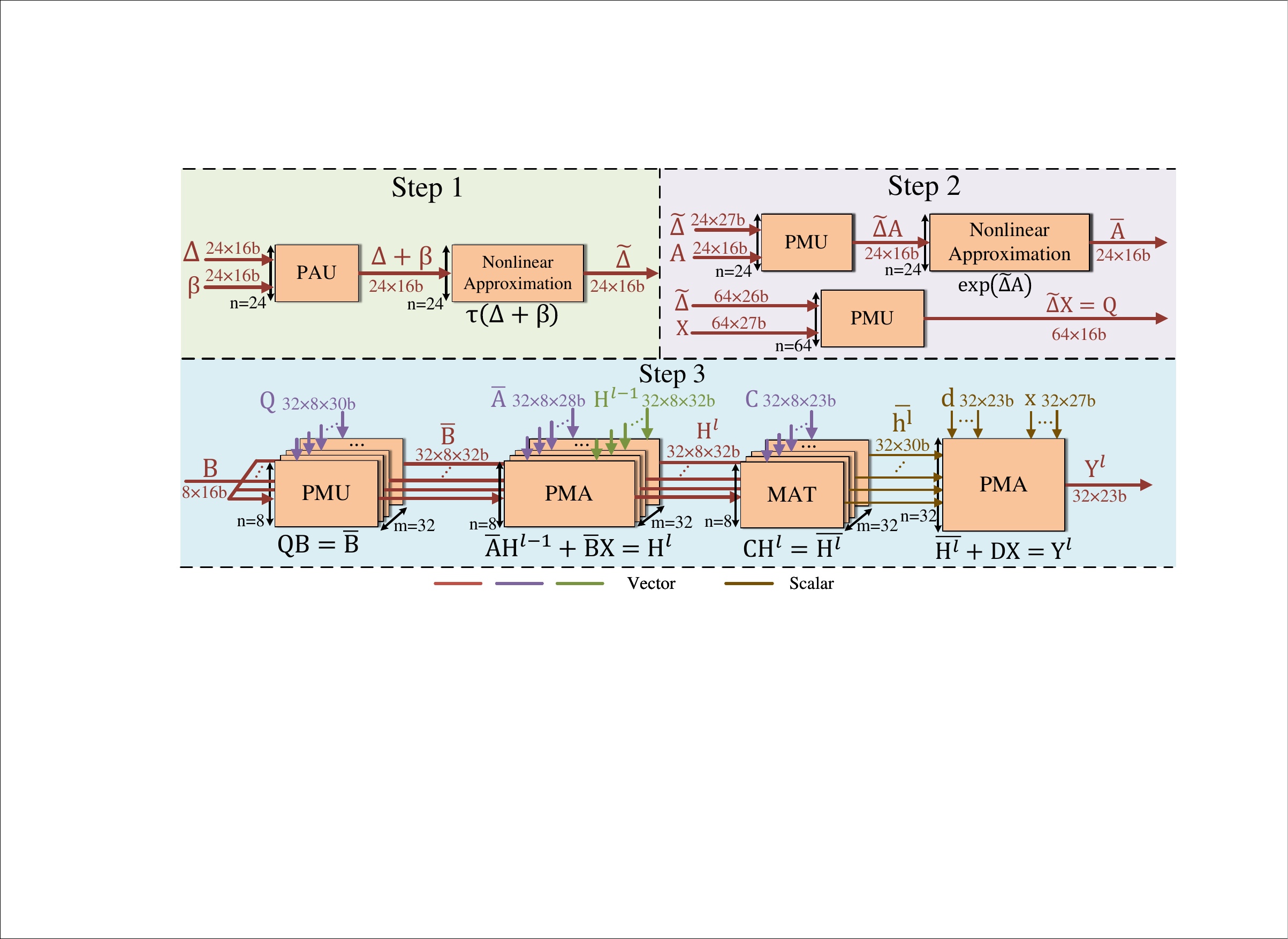}}
\caption{Architecture and computational flow of the SSM module (24$\times$16b indicates 24 16-bit data).}
\label{fig:ssm}
\vspace{-1.8em}
\end{figure}

Following the three-step computations corresponding to Fig. \ref{dataflow}, we design the SSM Module as shown in Fig. \ref{fig:ssm}. The specific details are as follows:

\textbf{Step1:} We design a PAU and a Nonlinear Approximation Unit, each having an input vector length of 24. In the $SoftPlus$ mode configuration of the Nonlinear Approximation Unit, it computes and outputs $\widetilde{\Delta}\in R^{1\times24}$.

\textbf{Step2:} One of the paths consists of a PMU and a Nonlinear Approximation Unit both with an input vector length of 24. When the Nonlinear Approximation Unit is set to the exponential mode, this design outputs $\overline{A}\in R^{1\times24}$. Besides, a PMU with an input vector length of 64 computes $Q\in R^{1\times64}$.

\textbf{Step3:} Initially, we set up 32-parallel PMU and PMA units to generate hidden-state vectors $H^l\in R^{32\times 8}$ for the current $l$. Then, the $H^l$ undergoes inner product processing with $C$ by 32-parallel MAT, outputting the scalars $\overline{h^l}$.
Finally, the scalars of $D$ ($d$), the scalars of $X$ ($x$) and $\overline{h^l}$ are fed into a 32-input PMA unit to perform linear computation generating the SSM output $Y^l$.

\subsection{Nonlinear Approximation Unit in SSM}
\begin{figure}[tbp]
\centerline{\includegraphics[width=0.5\textwidth]{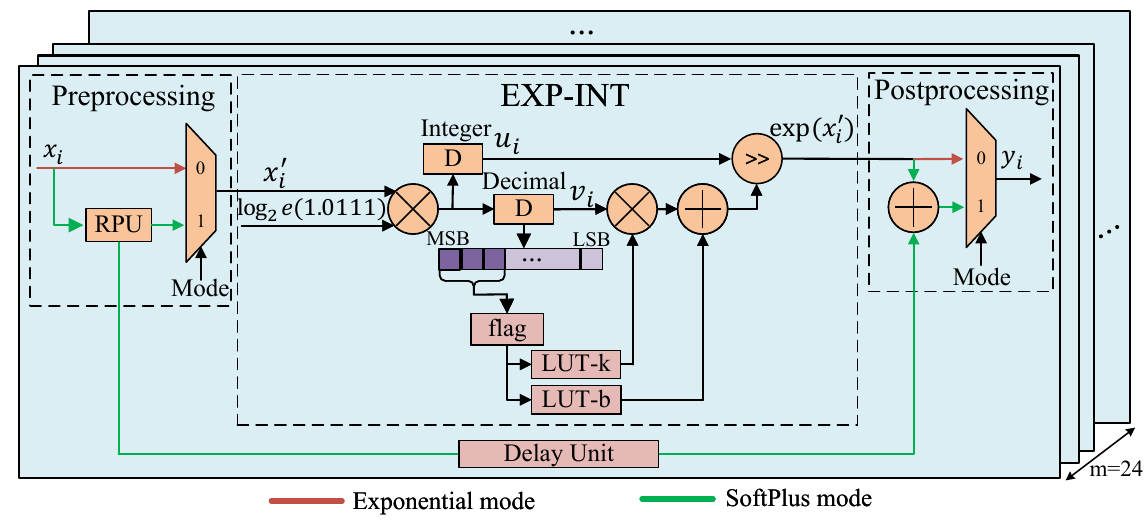}}
\caption{The multiplex structure of the Nonlinear Approximation Unit}
\label{fig:nonlinear}
\vspace{-1.8em}
\end{figure}

As depicted in Fig. \ref{fig:nonlinear}, we propose a 24-parallel Nonlinear Approximation Unit that can compute both the exponential function and $SoftPlus$ activation. It supports the 16-bit fixed-point vector with a length of 24 both inputs and outputs. The EXP-INT part fully implements the Eq. \eqref{func:exp}. When the input scalar $x_i\leq0$, the exponential mode is activated. 
In this mode, $x_i$ directly traverses the EXP-INT part to obtain the exponential output $exp(x_i)$. 
The exponential mode is also capable of computing the $SoftPlus$ when the $x_i\leq0$.

When $x_i>0$, the $SoftPlus$ mode is set. First, the Reverse Process Unit (RPU) in the Preprocessing part converts $x_i$ to its negative value $-x_i$, while $x_i$ is temporarily stored in the Delay Unit. Then, $-x_i$ passes through the EXP-INT part to compute the intermediate result $exp(-x_i)$. Finally, $exp(-x_i)$ and $x_i$ from the Delay Unit are summed by the adder in the Postprocessing part.
\section{Experimental Results}
\subsection{Experimental Setup}

\textbf{Models and Datasets.} During the prompt prefill stage, because of the limited GPU memory, we employ the Mamba2-130M model to evaluate the accuracy of our algorithm under 8-bit activation values and weights quantization (denoted as W8A8). We report the perplexity and zero-shot accuracy on 7 tasks: Lambada, Hellaswag, Piqa, Arc-easy, Arc-challenge, Winogrande, and Openbookqa \cite{b2} by using the lm-evaluation-harness \cite{b18} tool. Moreover, the Mamba2-130M is also utilized to evaluate the speedup ratio of FastMamba. In the decode stage, we use the Mamba2-2.7B model to evaluate the throughput and energy efficiency of FastMamba.

\textbf{Algorithm Evaluation Baselines.} We choose the Mamba2-130M model in the data format of half-precision floating-point (denoted as FP16) as the baseline. We also apply Normal Quantization (NormalQ) and Smooth Quantization (SmoothQ) \cite{b7} that perform W8A8 quantization only on linear layers as a comparison. In NormalQ, no optimization is processed for the outliers of activation values and weights.

\textbf{Hardware Evaluation Platforms.} We implement FastMamba on the Xilinx Virtex-7 VC709 FPGA. 
We selected the Intel Xeon
Silver 4210R CPU and the NVIDIA RTX 3090 GPU as baseline platforms, denoted as CPU and GPU, respectively. The system configuration of the CPU, GPU, and FPGA are presented in Table \ref{tab:hardware}. 

\begin{table}[htbp]
\centering
\caption{System Configuration.}  
\label{tab:hardware}
\begin{adjustbox}{width=0.5\textwidth}
\begin{tabular}{cccc}  
\Xhline{1pt}
\rule{0pt}{8pt}
&\textbf{CPU} & \textbf{GPU} & \textbf{FastMamba}\\
\hline  
\rule{0pt}{8pt}
\multirow{2}{*}{\textbf{Platform}}  & Intel Xeon & NVIDIA GeForce  & Xilinx Virtex-7 \\
& Silver 4210R(14nm) & RTX 3090(8nm) & VX690T(28nm) \\
\hline  
\rule{0pt}{8pt}
\textbf{Frequency} & 2400MHz & 1395MHz & 250MHz \\
\hline  
\rule{0pt}{8pt}
\textbf{Computing} & 10 & 328 & 3333 \\
\textbf{Units} & Cores & Tensor Cores & DSPs\\
\hline  
\rule{0pt}{8pt}
\textbf{Throughput} & - & 111 token/s & 5.68 token/s\\
\hline  
\rule{0pt}{8pt}
\textbf{Energy Efficiency} & - & 0.37 token/(s$\cdot$W) & 0.61 token/(s$\cdot$W)\\ 
\Xhline{1pt}
\end{tabular}
\end{adjustbox}
\vspace{-1.5em}
\end{table}

\subsection{Accuracy Evaluations}
Table \ref{tab:accruacy} presents the accuracy performance of W8A8 quantization using different methods based on Mamba2-130M. In the table, FastMamba-LQ only quantizes the linear layer, while FastMamba quantizes the linear layer, convolution layer, and SSM block.

The results show that FastMamba-LQ surpasses NormalQ and SmoothQ in both perplexity and accuracy. The outlier features with distinct degrees still exist in activation values and weights after NormalQ and SmoothQ processing. In contrast, the Hadamard transformation evenly disperses the outliers of activation values and weights across channels, significantly mitigating the impact of outlier features and attaining optimal accuracy. Due to applying the PoT quantization in the convolution layer and SSM block, FastMamba experiences a minor accuracy reduction within an acceptable range compared to FastMamba-LQ.

\begin{figure}[tp]
\centerline{\includegraphics[width=0.5\textwidth]{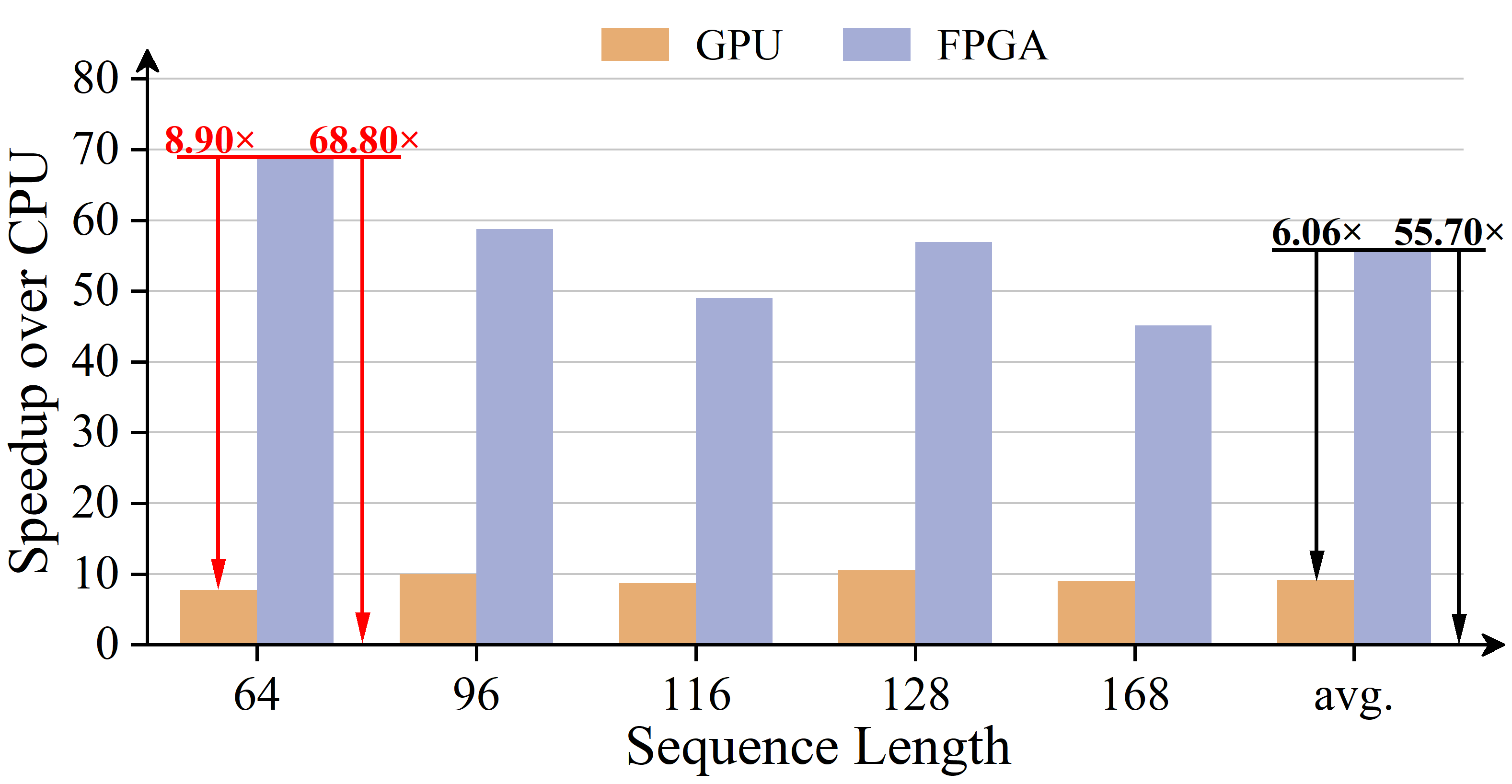}}
\caption{Comparison of speedup improvement to CPU and GPU on Mamba2-130M with different input sequence length during prompt prefill stage.}
\label{fig:speedup}
\vspace{-1em}
\end{figure}

\begin{figure}[!t]
\centerline{\includegraphics[width=0.5\textwidth]{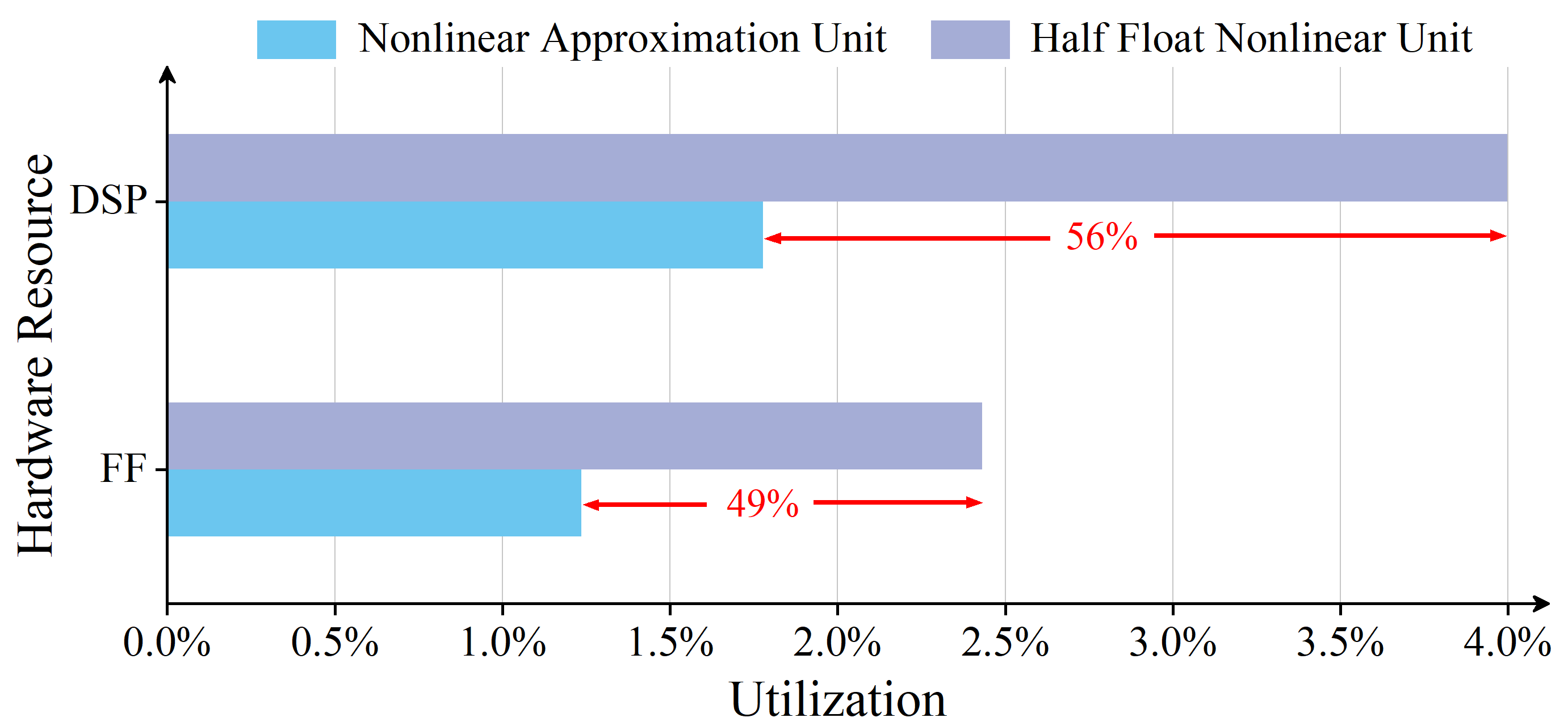}}
\caption{The hardware resource utilization rates of the Nonlinear Approximation Unit and the Half Float Nonlinear Unit.}
\label{fig:nonresource}
\vspace{-0.5em}
\end{figure}

\begin{table}[!t]
\centering
\caption{FastMamba FPGA Resources Utilization}  
\label{tab:resource}
\begin{adjustbox}{width=0.5\textwidth}
\begin{tabular}{ccccc}  
\Xhline{1pt}
\rule{0pt}{8pt}
\textbf{Component} & \textbf{LUT} & \textbf{FF} & \textbf{DSP} & \textbf{BRAM}  \\
\hline  
\rule{0pt}{8pt}
\textbf{Linear} & 132030(30.5\%) & 84514(9.8\%) & 48 (1.3\%) & 0 \\
\hline
\rule{0pt}{8pt}
\textbf{Convolution} & 14125(3.3\%) & 13201(1.5\%) & 256(7.1\%) & 0 \\
\hline
\rule{0pt}{8pt}
\textbf{SSM} & 73597(17.0\%) & 58196(6.7\%) & 2376(66.0\%) & 0 \\
\hline
\rule{0pt}{8pt}
\textbf{RMS Norm.} \& \textbf{SiLU}& 57315(13.2\%) & 87633(10.1\%) & 461(12.8\%) & 0 \\ 
\hline
\rule{0pt}{8pt}
\textbf{Buffer} & 13597(3.1\%) & 64898(7.5\%) & 0 & 956(65.0\%) \\
\hline
\rule{0pt}{8pt}
\textbf{Others} & 44120(10.2\%) & 46022(5.3\%) & 192(5.3\%) & 0 \\
\hline
\rule{0pt}{8pt}
\textbf{Total} & 334784(77.3\%) & 354464(40.9\%) & 3333(92.5\%) & 956(65.0\%) \\
\Xhline{1pt}
\end{tabular}
\end{adjustbox}
\vspace{-1.8em}
\end{table}

\subsection{Hardware Evaluations}
\subsubsection{Prompt Prefill Speedup} As shown in Fig. \ref{fig:speedup}, during the prompt prefill stage on Mamba2-130M, compared with CPU and GPU, FastMamba achieves a maximum speedup ratio of 68.80$\times$/8.90$\times$ and an average speedup ratio of 55.70$\times$/6.06$\times$ at the typical evaluation length. 
The improvement in acceleration performance mainly originates from fixed-point quantization, parallel VPUs, and pipeline design.

\subsubsection{Decode Throughput and Energy Efficiency} As Mamba2 does not support CPU decode inference, we only report the throughput and energy efficiency of the GPU and FastMamba on Mamba2-2.7B, as shown in Table \ref{tab:hardware}. FastMamba exhibits an energy efficiency improvement of 1.65$\times$ compared to the GPU baseline.

\subsubsection{Resource report} Table \ref{tab:resource} presents the resource consumption of FastMamba. The Hadamard-based linear module and the SSM module consume the majority of FPGA resources. In the Hadamard-based linear module, the 8-bit MAT units are mainly implemented using Look-Up Table (LUT) units. In the SSM module, as there are the most VPUs, it consumes the largest number of DSP resources. As illustrated in Fig. \ref{fig:nonresource}, our Nonlinear Approximation Unit saves 56\% of DSP resources and 49\% of flip-flop (FF) resources, compared with the Half Float Nonlinear Unit using FP16 data format.

\section{Conclusion}
This paper proposes FastMamba, a dedicated accelerator on FPGA for Mamba2 with hardware-algorithm co-design. 
We achieve 8-bit quantization through Hadamard transformation for the linear layer while applying PoT quantization for the convolution layer and SSM block.  
To optimize the nonlinear functions in the SSM block, we propose a hardware-friendly and efficient linear approximation and design a dedicated Nonlinear Approximation Unit.
Moreover, we implement an accelerator on Xilinx VC709 FPGA utilizing parallel vector processing units and high-efficiency pipelining. 
For the prefill task on Mamba2-130M, FastMamba achieves a maximum 68.80$\times$ and 8.90$\times$ speedup over the Intel Xeon 4210R CPU and NVIDIA RTX 3090 GPU, respectively. 
In the output decode experiment with Mamba2-2.7B, our design achieves a 1.65$\times$ higher energy efficiency than the RTX 3090 GPU. 



\end{document}